\documentclass[conference]{IEEEtran}

\usepackage{amsfonts}
\usepackage{amssymb}
\usepackage{cite}
\usepackage[cmex10]{amsmath}
\usepackage{float}
\usepackage{color}
\usepackage{stfloats,fancyhdr}
\usepackage{mathrsfs}
\usepackage{algorithm}
\usepackage{algorithmic}
\usepackage{multirow}
\usepackage{changepage}
\usepackage[normalem]{ulem}
\usepackage{geometry}
\usepackage{caption}
\usepackage{booktabs}
\usepackage{multirow}
\IEEEoverridecommandlockouts

\ifCLASSINFOpdf
\usepackage[pdftex]{graphicx}
\DeclareGraphicsExtensions{.pdf,.jpeg,.png}
\else
\usepackage[dvips]{graphicx}
\DeclareGraphicsExtensions{.eps}
\fi
\usepackage{subfigure}
\usepackage{fancybox,dashbox}
\begin{document}

\title{Antenna Selection for MIMO-NOMA Networks}

\author{Yuehua Yu$^\dag$, He Chen$^\dag$, Yonghui Li$^\dag$, Zhiguo Ding$^*$, and Branka Vucetic$^\dag$\\
$^\dag$School of Electrical and Information Engineering, The University of Sydney, Australia\\
$^*$School of Computing and Communications, Lancaster University, UK\\
Email: $^\dag$\{yuehua.yu, he.chen, yonghui.li, branka.vucetic\}@sydney.edu.au, $^*${z.~ding@lancaster.ac.uk}.}

\maketitle

\begin{abstract}

This paper considers the joint antenna selection (AS) problem for a classical two-user non-orthogonal multiple access (NOMA) network where both the base station and users are equipped with multiple antennas. Since the exhaustive-search-based optimal AS scheme is computationally prohibitive when the number of antennas is large, two computationally efficient joint AS algorithms, namely max-min-max AS (AIA-AS) and max-max-max AS (A$^3$-AS), are proposed to maximize the system sum-rate. The asymptotic closed-form expressions for the average sum-rates for both AIA-AS and A$^3$-AS are derived in the high signal-to-noise ratio (SNR) regime, respectively. Numerical results demonstrate that both AIA-AS and A$^3$-AS can yield significant performance gains over comparable schemes. Furthermore, AIA-AS can provide better user fairness, while the A$^3$-AS scheme can achieve the near-optimal sum-rate performance.
\end{abstract}


\IEEEpeerreviewmaketitle

\section{Introduction}
The non-orthogonal multiple access (NOMA) technique has been recently regarded as a promising solution to significantly improve the spectral efficiency of 5G wireless networks~\cite{Ref-Saito}. By superposing the information of multiple users in the power domain, multiple users can be served within the same time, frequency and code domain. Different from the conventional water-filling power allocation strategy, the NOMA technique allocates more transmit power to the users with poor channel conditions (i.e., \textit{weak users}). In this case, these users can decode their higher-power-level signals directly by treating others' signals as noise. In contrast, those users with better channel conditions (i.e., \textit{strong users}) adopt the successive interference cancellation (SIC) technique for signal detection. Specifically, the weak users' messages are first decoded and subtracted before the strong users' massages with lower-power-levels can be recovered. It has been demonstrated that both the system throughput and user fairness are significantly improved in NOMA systems compared to conventional orthogonal multiple access (OMA) systems~\cite{Ref-Saito}.

Recent works have attempted to apply the multiple-input-multiple-output (MIMO) techniques into NOMA systems (MIMO-NOMA) to exploit the spatial degrees of freedom~\cite{Ref-MIMO1,Ref-MIMO2,Ref-Massive-MIMO}. Although the capacity performance can potentially scale up with the increase of the number of antennas, this superior performance comes at the price of expensive RF chains at terminals. To avoid the heavy burden of hardware expenses while preserving the diversity benefits from MIMO, the antenna selection (AS) technique has been well recognized as an effective solution \cite{Ref-AS}. There are only a few papers that considered the AS problem for MIMO-NOMA systems in open literature. Specifically, a transmit AS (TAS) algorithm was proposed in \cite{Ref-TAS-in-NOMA}, and a joint TAS and user scheduling algorithm was studied in \cite{Ref-AS-in-Massvie-MIMO-NOMA}. However, both \cite{Ref-TAS-in-NOMA} and \cite{Ref-AS-in-Massvie-MIMO-NOMA} only focused on the TAS design at the base station side as each user was assumed to equipped with single antenna. Moreover, the insightful analytical expressions of the system performance have not been derived in~\cite{Ref-TAS-in-NOMA}~and~\cite{Ref-AS-in-Massvie-MIMO-NOMA}.

To our best knowledge, the joint AS at both the base station and users for MIMO-NOMA systems is still an open problem. Though there are some previous literatures which studied the joint AS schemes in conventional MIMO-OMA systems, they cannot be extended to MIMO-NOMA systems directly. This is because there are severe inter-user interferences in MIMO-NOMA systems while the signals are transmitted in an interference free manner in MIMO-OMA systems. A native way to find the global optimal solution of the problem may require an exhaustive search (ES) over all possible antenna combinations, whose complexity would become unacceptable as the numbers of antennas at both the base station and users become large. Motivated by this, in this paper we propose two computationally efficient joint AS algorithms for MIMO-NOMA systems, namely max-min-max AS (AIA-AS) and max-max-max AS (A$^3$-AS), to maximize the system sum-rate. Specifically, AIA-AS tries to improve the performance of the instantaneous weak user while the A$^3$-AS scheme maximizes the performance of the instantaneous strong user. The asymptotic closed-form expressions for the average rates for both AIA-AS and A$^3$-AS are derived for high signal-to-noise (SNR) scenario. All analytical results are validated by computer simulations and it is demonstrated that both the AIA-AS and A$^3$-AS can yield significant performance gains over comparable schemes. Furthermore, AIA-AS can provide better user fairness while the A$^3$-AS scheme can achieve the near-optimal sum-rate~performance.

\section{System Model}

 \begin{figure}
  \centering
  \includegraphics[width=0.4\textwidth]{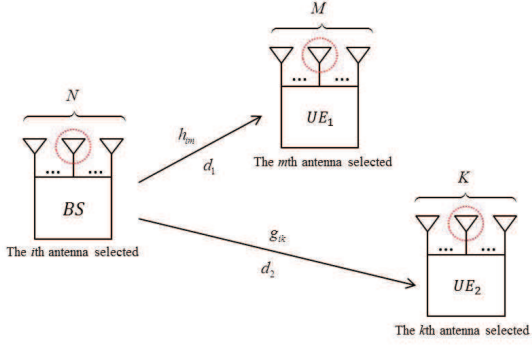}
  \caption{{Diagram of a MIMO-NOMA down-link scenario with multiple antennas at each node.}}
  \label{Fig:SystemModel}
\end{figure}

Consider a two-user MIMO-NOMA down-link scenario, since the two-user scenario is how NOMA is implemented in LTE-Advanced \cite{Ref-MIMO2}. As shown in Fig. \ref{Fig:SystemModel}, the base station (BS) is equipped with $N$ antennas, while user one (UE$_1$) and user two (UE$_2$) are equipped with $M$ and $K$ antennas, respectively. The channel matrix from the BS to UE$_1$ is denoted by ${\bf{H}} \in {\mathcal{C}^{N \times M}}$ and the one from the BS to UE$_2$ is denoted by ${\bf{G}} \in {\mathcal{C}^{N \times K}}$, where $\mathcal{C}^{m \times n}$ represents the set of all $m\times n$ matrices. We assume that the channels between the BS and users are spatially uncorrelated flat Raleigh fading, then the entries of $\mathbf{H}$ ($\mathbf{G}$), e.g., ${\tilde{h}_{im}}$ (${\tilde{g}_{ik}}$), can be modeled as independent and identically distributed (i.i.d.) complex Gaussian random variables, where $\tilde{h}_{im}$ ($\tilde{g}_{ik}$) represents the channel coefficient between the $i$th antenna of the BS and the $m$th ($k$th) antenna of UE$_1$ (UE$_2$). Specially, we define ${h}_{im}=|\tilde{h}_{im}|^2$ and ${g}_{ik}=|\tilde{g}_{ik}|^2$, where the operation $\left|{\cdot}\right|$ denotes the absolute value.

We assume all the nodes are limited with one RF chain based on the cost consideration. As such, in each resource block, the BS selects one (e.g., $i$th) out of $N$ available antennas to transmit information, while the users select one (e.g., $m$th and $k$th) out of $M$ and $K$ available antennas respectively to receive massages. To proceed, the global information of the channel amplitudes are assumed to be perfectly known at the~BS.

Let $\delta$ denote the indicator, defined as
\begin{eqnarray}\label{Equ_Delta}
\delta=\left\{ {\begin{array}{*{20}{c}}
{1,}&{{h}_{im}\ge {g}_{ik}},\\
{0,}&{{h}_{im}< {g}_{ik}}.
\end{array}} \right.
\end{eqnarray}

According to the principle of NOMA, the BS broadcasts the signals superposed in the power domain as
\begin{eqnarray}\label{Equ_x}
x=({1-\delta })(\sqrt{a}s_1 + \sqrt{b}s_2) + \delta(\sqrt{b}s_1 + \sqrt{a}s_2),
\end{eqnarray}where $s_i$ denotes the signal to UE$_i$ with $\mathbb{E}[{\left| {{s_i}} \right|^2}]=1$ and $\mathbb{E}[\cdot]$ denotes the expectation operation. $a$ and $b$ are the power allocation coefficients satisfying $a+b=1$. For notational simplicity, we assume that $a>b$ is set to guarantee that more power is allocated to the instantaneous weak~user.

The received signals at UEs are given by
\begin{eqnarray}\label{Equ_y}
{y_1}&=&\sqrt{P_s}\tilde{h}_{im}x+n_1,\\
{y_2}&=&\sqrt{P_s}\tilde{g}_{ik}x+n_2,
\end{eqnarray}where $P_s$ is the transmit power at the BS, and $n_i$ is the complex additive white Gaussian noise (AWGN) with variance $\sigma_i^2$. For simplicity, we assume $\sigma_1^2 = \sigma_2^2=\sigma^2$.

When $\delta=1$, UE$_2$ is the weak user and UE$_1$ is the strong user, hence the power level of $s_2$ is larger than that of $s_1$. In this case, UE$_2$ decodes $s_2$ directly by treating $s_1$ as noise. In contrast, UE$_1$ first decodes $s_2$ and subtracts it by performing SIC, then decodes its own $s_1$ without interference. For the case $\delta=0$, the decoding order is inverted. By using the fact that the channels are ordered, it can be easily verified that SIC can be carried out successfully, and the following two rates are achievable to the users:
\begin{eqnarray}\label{Equ_R1}
{R_1}\!\!\!&=&\!\!\!\left(1\!-\!\delta\right)\log_2\left(1\!+\!\frac{a{h}_{im}}{b{h}_{im}+\frac{1}{\rho}}\right)\!+\!\delta\log_2\left(1+\rho b{h}_{im}\right),\\
\label{Equ_R2}
{R_2}\!\!\!&=&\!\!\!\left(1\!-\!\delta\right)\log_2\left(1\!+\!\rho b{g}_{ik}\right)\!+\!\delta\log_2\left(1+\frac{a{g}_{ik}}{b{g}_{ik}+\frac{1}{\rho}}\right),
\end{eqnarray}where ${\rho} = {{{P_s}} \mathord{\left/ {\vphantom {{{P_s}} {{\sigma ^2}}}} \right.\kern-\nulldelimiterspace} {{\sigma ^2}}}$ is the transmit SNR. Accordingly, the achievable system sum-rate is given~by
\begin{eqnarray}\label{Equ_Rsum}
R_{\mathrm{sum}}&=&R_1 +R_2\nonumber\\
&=& \log_2\left(1+\rho b\tilde{\gamma}_i^{s}\right)+\log_2\left(1+\frac{a\tilde{\gamma}_i^{w}}{b\tilde{\gamma}_i^{w}+1/\rho}\right),
\end{eqnarray}where $\tilde{\gamma}_i^s=\max\left({h}_{im},{g}_{ik}\right)$ denotes the instantaneous channel gain of the strong user and $\tilde{\gamma}_i^w=\min\left({h}_{im},{g}_{ik}\right)$ denotes the instantaneous channel gain of the weak user.

In order to maximize the achievable system sum-rate, we need to solve the AS problem given by
\begin{eqnarray}\label{Equ_p1}
\mathbf{P1}:~\left\{ i^*,m^*,k^*\right\}  = \mathop {\arg \max }\limits_{i \in {\mathcal{N}},m \in {\mathcal{M}},k \in {\mathcal{K}}} {R_{\mathrm{sum}}}\left({h}_{im},{g}_{ik}\right),
\end{eqnarray}where $\mathcal{N}=\{1,2,\cdots,N\}$, $\mathcal{M}=\{1,2,\cdots,M\}$ and $\mathcal{K}=\{1,2,\cdots,K\}$.

It is straightforward to see that the optimization problem is an NP-hard problem, which means that the global optimal solution to the problem cannot be efficiently achieved. Finding the optimal combination of antennas at both the BS and users may require an exhaustive search with the complexity of $\mathcal{O}\left( {NMK} \right)$\footnote{The big $\mathcal{O}$ notion is usually used in the efficiency analysis of algorithms. $q(x)=\mathcal{O}(p(x))$ when $\lim\limits_{x\rightarrow\infty}|\frac{q(x)}{p(x)}|=c, 0<c<\infty$.}. This becomes unaffordable when $N$, $M$ and $K$ become large. Motivated by this, in next section we will develop two computationally efficient AS algorithms with dramatically reduced computational complexity.

\section{Proposed Antenna Selection Algorithms}
We can easily observe from (\ref{Equ_Rsum}) that the sum-rate $R_{\mathrm{sum}}$ is an increasing function of both $\tilde{\gamma}_i^w$ and $\tilde{\gamma}_i^s$. Based on this observation, in this section we develop two novel joint AS algorithms, termed AIA-AS and A$^3$-AS, for the considered MIMO-NOMA system. In particular, AIA-AS aims to maximize $\tilde{\gamma}_i^w$, while A$^3$-AS targets to maximize $\tilde{\gamma}_i^s$. The principles of these two proposed algorithms are elaborated in the following two subsections.
\subsection{Max-min-max antenna selection (AIA-AS)}

%

AIA-AS mainly consists of three stages as below.

\begin{itemize}
   \item {\textbf{Stage 1}}. Find out the largest elements ${h_{i}^{\max}}$ and ${g_{i}^{\max}}$ for each row of ${\bf{H}}$ and ${\bf{G}}$, respectively.
  \begin{eqnarray}\label{Equ_AIA_Stage11}
  {{h}_{i}^{\max}} &=& \max \left({h}_{i1},\cdots,{h}_{iM}\right),\\
  \label{Equ_AIA_Stage12}
  {{g}_{i}^{\max}} &=& \max \left({g}_{i1},\cdots,{g}_{iK}\right).
  \end{eqnarray}Then each pair $({h}_{i}^{\max},{g}_{i}^{\max})$ is treated as one AS candidate where $i\in{\mathcal{N}}$. The set of all $N$ pairs can be written as $\mathcal{S}^{(1)}=\left\{({h}_{1}^{\max},{g}_{1}^{\max}),\cdots,({h}_{N}^{\max},{g}_{N}^{\max})\right\}$.

  \item {\textbf{Stage 2}}. Find out the relatively \textit{smaller} element $\gamma_i^w$ in each pair $({h}_{i}^{\max},{g}_{i}^{\max})$. That is,
  \begin{eqnarray}\label{Equ_AIA_Stage2}
  \gamma_i^w= \min ({h}_{i}^{\max},{g}_{i}^{\max}),~~i\in{\mathcal{N}}.
  \end{eqnarray}The set of the $N$ smaller elements are denoted by $\mathcal{S}^{(2)}_{AIA}=\{\gamma_1^w,\cdots,\gamma_N^w\}$.

  \item {\textbf{Stage 3}}. Find out the largest element in $\mathcal{S}^{(2)}_{AIA}$, i.e.,
  \begin{eqnarray}\label{Equ_AIA_Stage3}
  \gamma^w= \max (\gamma_i^w),~~i\in{\mathcal{N}}.
  \end{eqnarray}
\end{itemize}

We use ($i^*_\mathrm{I}$, $k^*_\mathrm{I}$) to denote the original row and column indexes of $\gamma^w$ when it lies in $\bf G$. In this case, the $i^*_\mathrm{I}$th antenna at the BS and the $k^*_\mathrm{I}$th antenna at UE$_2$ are selected, respectively. Meanwhile, we use $m^*_\mathrm{I}$ to denote the original column index of $\gamma^s={h}_{i^*_\mathrm{I}}^{\max}$. Therefore the $m^*_\mathrm{I}$th antenna at UE$_1$ would be selected concurrently. For the case that $\gamma^w$ lies in $\bf H$, the selected antenna indexes can be obtained similarly. It is worth noting that $\gamma^w$ ($\gamma^s$) coming from UE$_1$ or UE$_2$ is not static, but varies based on the instantaneous channel conditions and the corresponding AS results.

The AIA-AS scheme is formally described in Algorithm~\ref{alg:AIA-AS}.

 \begin{algorithm}[htb]
  \caption{Max-min-max antenna selection (AIA-AS).}
  \label{alg:AIA-AS}
  \begin{algorithmic}[1] 
  \begin{small}
  \REQUIRE~~\\ 
  {The channel matrix of UE$_1$ and UE$_2$: $\bf H$ and $\bf G$;}\\
  {The antenna sets of the BS, UE1 and UE2: $\mathcal{N}$, $\mathcal{M}$ and $\mathcal{K}$.}
  \ENSURE ~~\\ 
  {$\{i^*_\mathrm{I},m^*_\mathrm{I},k^*_\mathrm{I}\}  = \mathop {\arg \max }\limits_{i \in \mathcal{N},m \in \mathcal{M},k \in \mathcal{K}} R_{\mathrm{sum}}( {h}_{im},{g}_{ik})$.}\\
  \textbf{Stage 1:}
  \STATE {Find out ${h_{i}^{\max}}$ and ${g_{i}^{\max}}$ according to (\ref{Equ_AIA_Stage11})-(\ref{Equ_AIA_Stage12});}\\
  \STATE {Obtain $\mathcal{S}^{(1)}=\left\{({h}_{1}^{\max},{g}_{1}^{\max}),\cdots,({h}_{N}^{\max},{g}_{N}^{\max})\right\}$.\\
  \textbf{Stage 2:}}
  \STATE {$\gamma_i^w= \min ({h_{i}^{\max }},{g_{i}^{\max }})$ where $({h_{i}^{\max }},{g_{i}^{\max }})\in \mathcal{S}^{(1)}$;}\\
  \STATE {Obtain $\mathcal{S}^{(2)}_{AIA}=\{\gamma_1^w,\cdots,\gamma_N^w\}$.\\
  \textbf{Stage 3:}}
  \STATE {$\gamma^w= \max (\gamma_{i}^w)$ where $\gamma_{i}^w\in \mathcal{S}^{(2)}_{AIA}$;}
  \STATE {$i^*_\mathrm{I}\leftarrow$ the row index of $\gamma^w$;}
  \STATE {$\gamma^s=\{h_{i^*_\mathrm{I}}^{\max},g_{i^*_\mathrm{I}}^{\max}\}-\{\gamma^w\}$, where $\gamma^s$ is in the same row of~$\gamma^w$;}
  \STATE {$m^*_\mathrm{I}$ and $k^*_\mathrm{I}\leftarrow$ the column indexes of $\gamma^w$ and $\gamma^s$.}
  \RETURN {$i^*_\mathrm{I},m^*_\mathrm{I},k^*_\mathrm{I}$.} 
  \end{small}
  \end{algorithmic}
  \end{algorithm}

\subsection{Max-max-max antenna selection (A$^3$-AS)}
Similar to the AIA-AS algorithm, A$^3$-AS scheme also has three stages. The main difference between A$^3$-AS and AIA-AS lies in the second stage. Specifically, for each pair of $({h_{i}^{\max }},{g_{i}^{\max }})$, A$^3$-AS selects the larger element to maximize the contribution of $\gamma^s$, while AIA-AS selects the smaller one to guarantee $\gamma^w$. Three stages of A$^3$-AS are elaborated as follow.

\begin{itemize}

  \item \textbf{Stage 1}. Similar to the stage 1 in AIA-AS, find out the set $\mathcal{S}^{(1)}=\{(h_{1}^{\max },g_{1}^{\max }),\cdots,(h_{N}^{\max },g_{N}^{\max })\}$.

  \item \textbf{Stage 2}. Find out the relatively \textit{larger} element in each pair of $(h_{i}^{\max },g_{i}^{\max })$. Mathematically, we have
  \begin{eqnarray}\label{Equ_A3AS_Stage_2}
  \gamma_{i}^s= \max ({h_{i}^{\max }},{g_{i}^{\max }}),~~i\in{\mathcal{N}}.
  \end{eqnarray}The set of the $N$ larger elements are denoted by $\mathcal{S}^{(2)}_{A^3}=\{\gamma_{1}^s,\cdots,\gamma_{N}^s\}$.

  \item \textbf{Stage 3}. Find out the largest element in $\mathcal{S}^{(2)}_{A^3}$, i.e.,
  \begin{eqnarray}\label{Equ_A3AS_Stage_3}
  \gamma^s= \max (\gamma_{i}^s),~~i\in{\mathcal{N}}.
  \end{eqnarray}
\end{itemize}

We use ($i^*_{\mathrm{A}}$, $m^*_{\mathrm{A}}$) to denote the original row and column indexes of $\gamma^s$ when it lies in $\bf H$. In this case, the $i^*_{\mathrm{A}}$th antenna at the BS and the $m^*_{\mathrm{A}}$th antenna at UE$_1$ are selected, respectively. Meanwhile, we use $k^*_{\mathrm{A}}$ to denote the original column index of $\gamma^w=g_{i^*_{\mathrm{A}}}^{\max}$. Therefore the $k^*_{\mathrm{A}}$th antenna at UE$_2$ would be selected simultaneously. For the case that $\gamma^s$ lies in $\bf G$, the selected antenna indexes can be obtained similarly.

The process of A$^3$AS is formally described in Algorithm \ref{alg:A3AS}.

  \begin{algorithm}[htb]
  \caption{Max-max-max antenna selection (A$^3$-AS).}
  \label{alg:A3AS}
  \begin{algorithmic}[1] 
  \begin{small}
  \REQUIRE ~~\\ 
  {The channel matrix of UE$_1$ and UE$_2$: $\bf H$ and $\bf G$;}\\
  {The antenna sets of the BS, UE$_1$ and UE$_2$: $\mathcal{N}$, $\mathcal{M}$ and $\mathcal{K}$.}
  \ENSURE ~~\\ 
  {$\{i^*_\mathrm{A},m^*_\mathrm{A},k^*_\mathrm{A}\}  = \mathop {\arg \max }\limits_{i \in {\mathcal{N}},m \in {\mathcal{M}},k \in {\mathcal{K}}} {R_{\mathrm{sum}}}({h}_{im},{g}_{ik})$.}\\
  \textbf{Stage 1:}
  \STATE {Similar to the stage one of A$^3$-AS in algorithm 1, obtain $\mathcal{S}^{(1)}=\{(h_{1}^{\max},g_{1}^{\max}),\cdots,(h_{N}^{\max},g_{N}^{\max})\}$.}\\
  \textbf{Stage 2:}
  \STATE {$\gamma_i^s= \max ({h_{i}^{\max}},{g_{i}^{\max}})$ where $({h_{i}^{\max}},{g_{i}^{\max}})\in \mathcal{S}^{(1)}$;}\\
  \STATE \small{Obtain $\mathcal{S}^{(2)}_{A^3}=\{\gamma_1^s,\cdots,\gamma_N^s\}$.}\\
  \textbf{Stage 3:}
  \STATE {$\gamma^s= \max (\gamma_i^s)$ where $\gamma_i^s\in \mathcal{S}^{(2)}_{A^3}$;}
  \STATE {$i^*_\mathrm{A}\leftarrow$ the row index of $\gamma^s$;}
  \STATE {$\gamma^w=\{h_{i^*_\mathrm{A}}^{\max},g_{i^*_\mathrm{A}}^{\max}\}-\{\gamma^s\}$, where $\gamma^w$ is in the same row of~$\gamma^s$;}
  \STATE {$m^*_\mathrm{A}$ and $k^*_\mathrm{A}\leftarrow$ the column indexes of $\gamma^s$ and $\gamma^w$.}
  \RETURN {$i^*_\mathrm{A},m^*_\mathrm{A},k^*_\mathrm{A}$. }
  \end{small}
  \end{algorithmic}
  \end{algorithm}

\subsection{User fairness}
To evaluate the user fairness of the proposed two AS algorithms in MIMO-NOMA system, the Jain's fairness index \cite{Ref-Fairness} is adopted in this paper. Specifically, the Jain's fairness index for the aforementioned two-users scenario can be expressed as
\begin{eqnarray}\label{Equ_Fairness}
\eta=\frac{(R_1+R_2)^2}{2(R_1^2+R_2^2)}.
\end{eqnarray}Jain's fairness index is bounded between 0 and 1 with the maximum achieved by equaling user's rates.

\subsection{Computational complexities}
As mentioned before, the complexity of the optimal selection algorithm achieved by the exhaustive search is as high as $\mathcal{O}\left(NMK\right)$. In other words, the exhaustive search needs to calculate the achievable rate for all the $NMK$ combinations before finding out the optimal antenna triple. When the number of antennas at each node becomes large, the computational burden would increase significantly.

In contrast, both the two proposed AS algorithms dramatically reduce the selection complexity to $\mathcal{O}\big(N(M+K+3)\big)$, where the main computation only lies in sorting the channel gains. For the case $N=M=K$, we can find that the complexity of AIA-AS and A$^3$-AS is approximately $\mathcal{O}(N^2)$, which reduces an order of magnitude compared to the complexity of $\mathcal{O}(N^3)$ for the optimal ES scheme.

\section{Performance Analysis of Proposed Algorithms}
In this section, we derive asymptotic close-form expressions for the system average sum-rates of AIA-AS and A$^3$-AS algorithms for the high SNR scenario.

Assuming the flat Raleigh fading channel, ${h}_{im}=|\tilde{h}_{im}|^2\geq 0$ is then an exponentially distributed random variable with the distribution given by
\begin{eqnarray}\label{Equ_h_im_cdf_pdf}
F_{{h}}(x)=1-e^{-\Omega_h x},~f_{{h}}(x)={\Omega_h}{e^{-\Omega_h x}},~x\geq0,
\end{eqnarray}where $\Omega_h=1/\mathbb{E}[h_{im}]$, and $F(x)$ and $f(x)$ denote the cumulative distribution function (CDF) and the probability distribution function (PDF), respectively. Similarly, let $\Omega_g=1/\mathbb{E}[g_{ik}]$ and for any element in $\bf G$, e.g., $g_{ik}$, we have the CDF and PDF of ${g}_{ik}$ as~follow
\begin{eqnarray}\label{Equ_g_ik_cdf_pdf}
F_{{g}}(x)=1-e^{-\Omega_gx},~f_{{g}}(x)=\Omega_g{e^{-\Omega_gx}},~x\geq0.
\end{eqnarray}

Recall that the first stage of both AIA-AS and A$^3$-AS is to find out $h_{i}^{\max}$ and $g_{i}^{\max}$ for $i\in \mathcal{N}$ according to (\ref{Equ_AIA_Stage11})-(\ref{Equ_AIA_Stage12}). Therefore, we can obtain the distribution of $h_{i}^{\max}$ for $x\geq0$ as follow
\begin{eqnarray}\label{Equ_AIA-himax-cdf-pdf}
F_{h_{i}^{\max}}(x)&\!\!=\!\!&\left(1\!\!-\!\!e^{-\Omega_h x}\right)^{M}\overset{(c_1)}{=}\sum\nolimits_{i=0}^M\lambda_{i,M}e^{-i\Omega_hx},\\
f_{h_{i}^{\max}}(x)&\!\!=\!\!&-\sum\nolimits_{i=1}^Mi\Omega_h\lambda_{i,M}e^{-i\Omega_hx},
\end{eqnarray}where $\lambda_{i,M}=(-1)^i\binom{M}{i}$ and the expansion step $(c_1)$ is conducted based on the Binomial theorem.

Similarly, the CDF and PDF of $g_{i}^{\max}$ for $x\geq0$ are given~by
\begin{eqnarray}\label{Equ_AIA-gimax-cdf-pdf}
F_{g_{i}^{\max}}(x)&\!\!=\!\!&\left(1\!\!-\!\!e^{-\Omega_g x}\right)^{K}=\sum\nolimits_{j=0}^K\lambda_{j,K}e^{-j\Omega_gx},\\
f_{g_{i}^{\max}}(x)&\!\!=\!\!&-\sum\nolimits_{j=1}^Kj\Omega_g\lambda_{j,K}e^{-j\Omega_gx}.
\end{eqnarray}

Then the asymptotic analysis for the sum-rates of AIA-AS and A$^3$-AS will be obtained in the following subsections,~respectively.

\subsection{Analytical sum-rate of the AIA-AS algorithm.}
In the second stage of AIA-AS, it is to find out the relatively smaller element $\gamma_{i}^{w}=\min(h_{i}^{\max},g_{i}^{\max})$ in each row. Thus, the CDF of $\gamma_{i}^{w}$ for $x\geq0$ can be calculated as follows:
\begin{eqnarray}\label{Equ_AIA_gamma_w_i_CDF}
F_{\gamma_i^{w}}(x)&=&\mathrm{Pr}\left\{\min(h_{i}^{\max},g_{i}^{\max})<x\right\}\nonumber\\
&=&1-\sum_{i=1}^{M}\sum_{j=1}^{K}\lambda_{i,M}\lambda_{j,K}e^{-(i\Omega_h+j\Omega_g)x}.
\end{eqnarray}

In the third stage of AIA-AS, it is to find out $\gamma^w=\max(\gamma_i^w)$ for $i\in\mathcal{N}$. In case $\gamma^w$ lies in the $i^*_\mathrm{I}$th row, we first define $\hat{\gamma}^w=\max\limits_{i\neq i^*_\mathrm{I}}(\gamma_i^w)$ and obtain the CDF of $\hat{\gamma}^w$ as follows:
\begin{eqnarray}\label{Equ_AIA_gamma_w_uneqnal_CDF}
F_{\hat{\gamma}^w}(x)=[F_{\gamma_i^{w}}(x)]^{N-1}\overset{(c2)}{=}\sum_{\ell}C_\ell{t_\ell e^{-\xi_\ell x}},
\end{eqnarray}where the step $(c2)$ is expanded according to the Multinomial theorem. Specifically, $\ell_0+\cdots+\ell_{MK}=N-1$, the multinomial coefficient $C_\ell=\binom{N-1}{\ell_0,\cdots,\ell_{MK}}=\frac{(N-1)!}{\ell_0!\cdots\ell_{MK}!}$, $t_{\ell}=\prod_{\substack{1\leq i\leq M\\1\leq j\leq K}} (-\lambda_{i,M}\lambda_{j,K})^{\ell_{ij}}$ and $\xi_{\ell}=\sum_{i=1}^M\sum_{j=1}^K(i\Omega_h+j\Omega_j)\ell_{ij}$.

Next we need to obtain the CDF and PDF of $\gamma^s=\max({h_{i^*_\mathrm{I}}^{\max}},{g_{i^*_\mathrm{I}}^{\max}})$ which lies in the same $i^*_\mathrm{I}$th row with $\gamma^w$. By applying some algebraic manipulations, we have
\begin{eqnarray}\label{Equ_AIA_gamma_s_CDF}
F_{\gamma^s}(x)&\!\!=\!\!&\mathrm{Pr}\left\{\max(h_{i^*_\mathrm{I}}^{\max},g_{i^*_\mathrm{I}}^{\max})<x, \gamma_{i^*_\mathrm{I}}^w\geq \hat{\gamma}^w \right\}\nonumber\\
&\!\!=\!\!&N\mathrm{Pr}(\hat{\gamma}^w <g_{i^*_\mathrm{I}}^{\max}<h_{i^*_\mathrm{I}}^{\max}<x)\nonumber\\
&\!\!+\!\!&N\mathrm{Pr}(\hat{\gamma}^w <h_{i^*_\mathrm{I}}^{\max}<g_{i^*_\mathrm{I}}^{\max}<x),
\end{eqnarray}and
\begin{eqnarray}\label{Equ_AIA_Kappa_n_PDF}
f_{\gamma^s}(x)&=&Nf_{h_i}^{\max}(x)\int_0^xf_{g_i}^{\max}(y)\int_0^yf_{\hat{\gamma}^w}(z)\mathrm{d}z\mathrm{d}y\nonumber\\
&+&Nf_{g_i}^{\max}(x)\int_0^xf_{h_i}^{\max}(y)\int_0^yf_{\hat{\gamma}^w}(z)\mathrm{d}z\mathrm{d}y\\
&=&\sum_{i=1}^M\sum_{j=1}^K\sum_{\ell}C_\ell t_\ell \zeta_{ij}\big(\psi(i\Omega_h,j\Omega_g)+\psi(j\Omega_g,i\Omega_h)\big),\nonumber
\end{eqnarray}where $\psi(\mu_1,\mu_2)=e^{-\mu_1x}\left(\frac{e^{-\mu_2 x}-1}{\mu_2}-\frac{e^{-(\mu_2+\xi_\ell) x}-1}{\mu_2+\xi_\ell}\right)$ and $\zeta_{ij}=Nij\Omega_h\Omega_g\lambda_{i,M}\lambda_{j,K}$.

By observing (\ref{Equ_Rsum}), we can approximate the achievable rate of the instantaneous weak user as a constant in the high SNR scenario, i.e. $\log_2\left(1+\frac{a\tilde{\gamma}_i^{w}}{b\tilde{\gamma}_i^{w}+1/\rho}\right)\overset{\rho\rightarrow \infty}{\approx}{\log_2\left(1/b\right)}$. In this case, we can find the approximation of the system average sum-rate as follow
\begin{eqnarray}\label{Equ_AIA_sum}
\bar{R}_{\mathrm{sum}}^{\mathrm{AIA}}&\approx&\int_0^\infty\log_2(1+b\rho x)f_{\gamma^s}(x)\mathrm{d}x+\log_2\frac{1}{b}\\
&=&\log_2\frac{1}{b}+\sum_{i=1}^M\sum_{j=1}^K\sum_{\ell}\frac{C_\ell t_\ell}{\ln2}\left(T_1+T_2+T_3+T_4\right),\nonumber
\end{eqnarray}where $T_n,~n=\{1,2,3,4\}$ are given by
\begin{eqnarray}
T_1&=&-\int_0^\infty\frac{\xi_{\ell}\zeta_{ij}\ln(1+b\rho x)e^{-j\Omega_gx}}{i\Omega_h\phi_i}\mathrm{d}x=\frac{\xi_\ell\tilde{\zeta}_{ij}}{\phi_i}\chi(j\Omega_g),\nonumber\\
T_2&=&-\int_0^\infty\frac{\xi_{\ell}\zeta_{ij}\ln(1+b\rho x)e^{-i\Omega_hx}}{j\Omega_g\phi_j}\mathrm{d}x=\frac{\xi_\ell\tilde{\zeta}_{ij}}{\phi_j}\chi(i\Omega_h),\nonumber\\
T_3&=&-\int_0^\infty\frac{\zeta_{ij}\phi_{ij,2}\ln(1+b\rho x)e^{-\phi_{ij,1}x}}{\phi_i\phi_j}\mathrm{d}x=\frac{{\zeta}_{ij}\phi_{ij,2}\chi(\phi_{ij,1})}{\phi_i\phi_j\phi_{ij,1}},\nonumber\\
T_4&=&\int_0^\infty\frac{\zeta_{ij}(i\Omega_h+j\Omega_g)\ln(1+b\rho x)e^{-(i\Omega_h+j\Omega_g)x}}{ij\Omega_h\Omega_g}\mathrm{d}x\nonumber\\
&=&-\tilde{\zeta}_{ij}\chi(i\Omega_h+j\Omega_g),\nonumber
\end{eqnarray}in which,
\begin{eqnarray}\label{Equ_AIA_notation}
\tilde{\zeta}_{ij}&=&N\lambda_{i,M}\lambda_{j,K},\nonumber\\
\phi_i&=&i\Omega_h+\xi_\ell,~~~\phi_j=j\Omega_g+\xi_\ell,\nonumber\\
\phi_{ij,1}&=&i\Omega_h+j\Omega_g+\xi_\ell,~~~\phi_{ij,2}=i\Omega_h+j\Omega_g+2\xi_\ell,\nonumber\\
\chi(x)&=&e^{\frac{x}{b\rho}}\mathrm{Ei}\left(-\frac{x}{b\rho}\right),\nonumber
\end{eqnarray}$\mathrm{Ei}(x)$ is the Exponential integral function and the integral of $T_n$ is obtained with the help of \cite[Eq. (4.337.2)]{Ref-Book-integrals}.

\subsection{Analytical sum-rate of the A$^3$-AS algorithm.}
Recall that in the A$^3$-AS algorithm, $\gamma^s$ is actually the larger element of $\max\limits_i(h_i^{\max})$ and $\max\limits_i(g_i^{\max})$. Here we denote $h^{\max}=\max\limits_i(h_i^{\max})$ and $g^{\max}=\max\limits_i(g_i^{\max})$ and obtain the corresponding distributions as follows:
\begin{eqnarray}\label{Equ_A3_hmax_gmax}
F_{h^{\max}}(x)&=&\sum\nolimits_{i=0}^{NM}\lambda_{i,NM}e^{-i\Omega_hx},\\
F_{g^{\max}}(x)&=&\sum\nolimits_{j=0}^{NK}\lambda_{j,NK}e^{-j\Omega_gx},\\
f_{h^{\max}}(x)&=&-\sum\nolimits_{i=1}^{NM}i\Omega_h\lambda_{i,NM}e^{-i\Omega_hx},\\
f_{g^{\max}}(x)&=&-\sum\nolimits_{j=1}^{NK}j\Omega_g\lambda_{i,NK}e^{-j\Omega_gx}.
\end{eqnarray}Then the CDF and PDF of $\gamma^s=\max(h^{\max},g^{\max})$ is given~by
\begin{eqnarray}\label{Equ_A3_kappa_n}
F_{\gamma^s}(x)&=&\mathrm{Pr}\left\{\max(h^{\max},g^{\max})<x\right\}\\
&=&\mathrm{Pr}\left(h^{\max}<g^{\max}<x)+\mathrm{Pr}(g^{\max}<h^{\max}<x\right),\nonumber\\
f_{\gamma^s}(x)&=&\int_0^x\big(f_g^{\max}(x)f_h^{\max}(y)+f_h^{\max}(x)f_g^{\max}(y)\big)\mathrm{d}y\nonumber\\
&=&\sum_{i=1}^{NM}\sum_{j=1}^{NK}\lambda_{i,NM}\lambda_{j,NK}\big(i\Omega_he^{-i\Omega_hx}+j\Omega_ge^{-j\Omega_gx}\nonumber\\
&-&(i\Omega_h+j\Omega_g)e^{-(i\Omega_h+j\Omega_g)x}\big).
\end{eqnarray}

Similarly, when $\rho\rightarrow\infty$, we can attain the asymptotic closed-form expression for the average sum-rate for the A$^3$-AS algorithm as follows
\begin{eqnarray}\label{Equ_A3_sum}
\bar{R}_{\mathrm{sum}}^{\mathrm{A}^3}&\approx&\int_0^\infty\log_2(1+b\rho_sx)f_{\gamma^s}(x)\mathrm{d}x+\log_2\frac{1}{b}\nonumber\\
&=&\log_2\frac{1}{b}+\frac{1}{\ln2}\sum_{i=1}^{NM}\sum_{j=1}^{NK}\lambda_{i,NM}\lambda_{j,NK}\nonumber\\
&\times&\big(\chi(i\Omega_h+j\Omega_g)-\chi(i\Omega_h)-\chi(j\Omega_g)\big),
\end{eqnarray}where $\chi(x)=e^{\frac{x}{b\rho}}\mathrm{Ei}\left(-\frac{x}{b\rho}\right)$ as in AIA-AS.
\section{Numerical Studies}

In this section, the performance of the proposed AS algorithms for MIMO-NOMA systems, i.e., AIA-AS and A$^3$-AS, is evaluated by using computer simulations. In all simulation, we set $M=K=2$, $\Omega_h=d_1^\alpha$, $\Omega_g=d_2^\alpha$, where $\alpha$ is the path-loss exponent and $\alpha=3$, $d_1$ ($d_2$) is the distance between the BS and UE$_1$ (UE$_2$).

\begin{figure}
  \centering
  \includegraphics[width=0.45\textwidth]{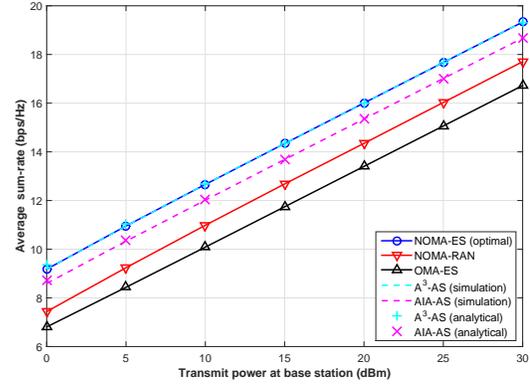}
  \caption{{Average sum-rate vs. transmit power, $N=2, d_1=30\mathrm{m}, d_2=100\mathrm{m}, a=0.6, b=0.4, \sigma=-70\mathrm{dBm}$.}}
  \label{Fig:Rate_Ps}
\end{figure}

Fig.~\ref{Fig:Rate_Ps} illustrates how the transmit power $P_s$ at the BS affects the system average sum-rate $\bar{R}_{\mathrm{sum}}$. As can be observed from Fig.~\ref{Fig:Rate_Ps}, when $P_s$ increases, $\bar{R}_{\mathrm{sum}}$ increases for all the schemes. Moreover, the performance of the proposed AIA-AS and A$^3$-AS schemes are much better than that of the random AS in NOMA scenarios (NOMA-RAN), since both AIA-AS and A$^3$-AS utilize the benefit brought by the multiple antennas settings at each node. Furthermore, the A$^3$-AS scheme can achieve the same performance as that of the optimal ES scheme in NOMA scenarios (NOMA-ES) but with much lower computational complexity. We should note that the analytical results match the simulation results for both AIA-AS and A$^3$-AS, which validates our theoretical analysis in Sec. IV. It is also worth pointing out that all the NOMA schemes outperform the ES scheme in OMA system (OMA-ES) over the entire region.

\begin{figure}
  \centering
  \includegraphics[width=0.45\textwidth]{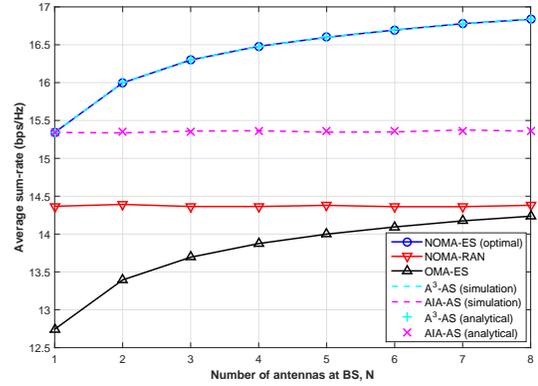}
  \caption{{Average sum-rate vs. $N$, $d_1=30\mathrm{m}, d_2=100\mathrm{m}, a=0.6, b=0.4, \sigma=-70\mathrm{dBm}, P_s=20\mathrm{dBm}$.}}
  \label{Fig:Rate_N}
\end{figure}

Fig.~\ref{Fig:Rate_N} illustrates how the number of antennas $N$ at the BS influences the average sum-rate $\bar{R}_{\mathrm{sum}}$. We can see from this figure that the sum-rates of the NOMA-RAN and AIA-AS keep constant when $N$ increases. For NOMA-RAN scheme, this is because it does not properly utilize the multiple antenna setting but selects one antenna at each node randomly. The reason for AIA-AS is that it guarantees the performance of the user with the poor channel gain $\gamma^w$, but not the user with the better channel condition $\gamma^s$, which contributes the most to $\bar{R}_{\mathrm{sum}}$. In contrast, the average sum-rate of A$^3$-AS increases along with $N$ and A$^3$-AS achieves the same performance as that of the optimal scheme. Again, all the NOMA schemes outperform the OMA-ES scheme in the entire region.

\begin{figure}
  \centering
  \includegraphics[width=0.45\textwidth]{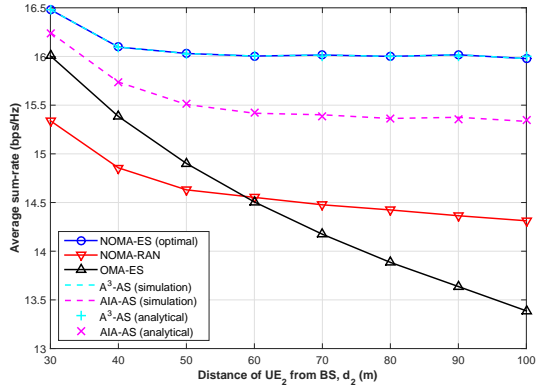}
  \caption{{Average sum-rate vs. $d_2$, $N=2, d_1=30\mathrm{m}, a=0.6, b=0.4, \sigma=-70\mathrm{dBm}, P_s=20\mathrm{dBm}$.}}
  \label{Fig:Rate_d}
\end{figure}

Fig.~\ref{Fig:Rate_d} depicts how the distance between the BS and users influences $\bar{R}_{\mathrm{sum}}$ for various AS schemes. Take a constant $d_1$ and a variable $d_2$ for example. We can observe that when $d_2$ increases, $\bar{R}_{\mathrm{sum}}$ decreases for all the schemes. We also note that both AIA-AS and A$^3$-AS outperform the NOMA-RAN and the OMA-ES schemes, and again A$^3$-AS achieves the same performance as NOMA-ES. Specially, there is a crossing between the curves for NOMA-RAN and OMA-ES. The reason for this is that in OMA-ES, when $d_2$ is much larger than $d_1$, the energy and frequency resources exclusively allocated to UE$_2$ are wasted since they contribute very little to~$\bar{R}_{\mathrm{sum}}$.

\begin{figure}
  \centering
  \includegraphics[width=0.45\textwidth]{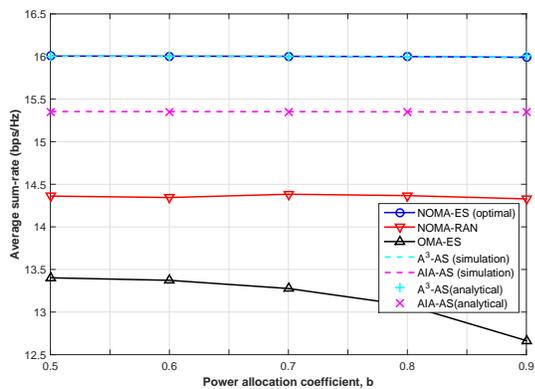}
  \caption{{Average sum-rate vs. b, $N=2, d_1=30\mathrm{m}, d_2=100\mathrm{m}, a=1-b, \sigma=-70\mathrm{dBm}, P_s=20\mathrm{dBm}$.}}
  \label{Fig:Rate_b}
\end{figure}

Fig.~\ref{Fig:Rate_b} demonstrates how the power allocation coefficient $b$ affects the $\bar{R}_{\mathrm{sum}}$ for various AS schemes. Interestingly we can see that all the NOMA schemes keep almost constant when $b$ increases. The main reason is that $\bar{R}_{\mathrm{sum}}\approx\log_2(\gamma^s\rho)$ when $\rho\rightarrow\infty$ and it is not affected by the value of $b$. In contrast, the performance of the OMA-ES scheme decreases when $b$ increases as more power are allocated exclusively to the user with the poor channel condition $\gamma^w$ which contributes little to~ $\bar{R}_{\mathrm{sum}}$.

\begin{figure}
  \centering
  \includegraphics[width=0.45\textwidth]{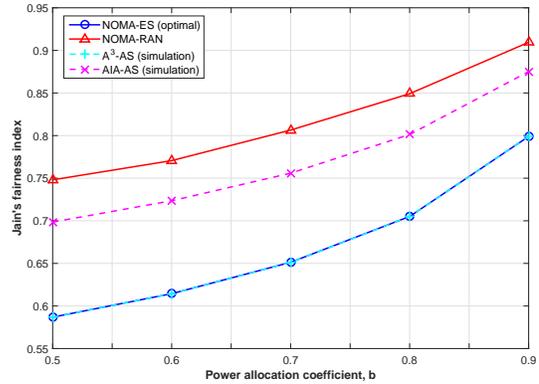}
  \caption{{Jain's fairness index vs. b, $N=4, d_1=60\mathrm{m}, d_2=100\mathrm{m}, a=1-b, \sigma=-70\mathrm{dBm}, P_s=20\mathrm{dBm}$.}}
  \label{Fig:fairness}
\end{figure}

Although the system sum-rate performance of A$^3$-AS is slightly better than that of AIA-AS, regarding the fairness between UE$_1$ and UE$_2$, we can observe in Fig.~\ref{Fig:fairness} that AIA-AS can provide better fairness than A$^3$-AS. In other words, in practice AIA-AS would be a better choice to balance the tradeoff between the system sum-rate and user fairness.

\section{Conclusion}
This paper studied the joint AS problem in a two-user MIMO-NOMA system. Two computationally efficient algorithms, i.e., AIA-AS and A$^3$-AS, were proposed to maximize the system sum-rate. The asymptotic closed-form expressions for the average sum-rates for both the proposed schemes were provided. Numerical simulations demonstrated that both AIA-AS and A$^3$-AS yield significant performance gains over the OMA-ES and NOMA-RAN schemes. Furthermore, AIA-AS provides better user fairness while the A$^3$-AS achieves the near-optimal sum-rate performance.






%

\end{document}